\documentclass[a4paper,11pt]{article}
\usepackage{jheppub} 
\usepackage{lineno}
\usepackage{graphicx}
\usepackage{float}
\usepackage{color}
\usepackage{amssymb,mathrsfs}
\usepackage{cases}
\usepackage{multirow}
\usepackage{subfigure}
\usepackage{graphicx}
\usepackage{epstopdf}
\def\td {\textmd{d}}

\def\e {\textmd{e}}

\newtheorem{theorem}{Theorem}
\newtheorem{lemma}{Lemma}

\title{Inverse problem of correlation functions in holography}
\author[a]{Bo-Wen Fan,}
\author[a]{Run-Qiu Yang}
\emailAdd{aqiu@tju.edu.cn}
\affiliation[a]{Center for Joint Quantum Studies and Department of Physics, School of Science, Tianjin University,\\ Yaguan Road 135, Jinnan District, 300350 Tianjin, P.~R.~China}

\abstract{This paper shows that the bulk metric of a planar/spherically/hyperbolically symmetric asymptotically anti-de Sitter static black brane/hole can be reconstructed from its boundary frequency 2-point correlation functions of two probe scalar operators by solving Gel'fand-Levitan-Marchenko integral equation. Since the frequency correlation function is easily handled in experiments and theories, this paper not only proposes a new method to ``measure'' the corresponding holographic spacetime for a material that has holographic dual but also provides an approach to experimentally check if a system has holographic dual.}
\begin{document}
\maketitle
\flushbottom
\noindent
\section{Introduction}
The mathematical expression of a physical law is usually a set of equations determined by the parameters, and its solutions stand for the results which can be compared with observations. Deriving the ``computed result'' for a given set of parameters is called ``solving the direct problem''. Conversely, obtaining the set of the parameters
from a given set of results is called ``solving the inverse problem''. In the study of holography duality, the direct problems are finding the boundary observables for the given bulk geometry and theory. Well, the inverse problems then try to find the bulk geometry and theory by using the boundary observables, which has a special name called ``bulk reconstruction'' in holography. The ``Solving the inverse problem'' plays essential roles both in theoretical studies and in practice of applied holography~\cite{Maldacena:1997re,Gubser:1998bc,Witten:1998qj,Witten:1998zw,Zaanen2015,Baggioli:2019rrs}. In theoretical viewpoint, bulk reconstruction is the central question in understanding the emerging spacetime in holography and answers how the gravitational degrees of freedom emerge from field theory.
The concept of bulk reconstruction likely originates from the work of Hamilton, Kabat, Lifschytz, and Lowe, who developed a method for reconstructing bulk operators~\cite{Hamilton:2006fh}. In terms of practical application, it will help us to build holographic spacetime and theory more efficiently by directly using the observable data when we try to use holographic duality to deal with strongly coupling materials.

A few advances have been pushed forward towards this issue. When conformal dimension $\Delta$ is much larger than spacetime dimensional $d$, boundary 2-point correlation functions can be expressed in terms of geodesics~\cite{Balasubramanian:1999zv}, and therefore have been used to reconstruct the bulk geometry~\cite{Hubeny:2012ry}. However, the requirement $\Delta\gg d$ restricts its application in practice since materials in general will not satisfy this requirement. There is also a method called ``bulk-cone singularities'', which uses the relationship between singular correlators and bulk null geodesics to recover some special spherically symmetric static bulk geometries~\cite{Bilson:2008ab,Hammersley:2006cp}.
In most situations, the materials we are concerned about have planar structures. For those boundary systems, this method cannot be applied since the bulk null geodesics of planar symmetric static  anti-de Sitter (AdS) black brane will no longer return to boundary again if they start from the boundary. What's more, locating all singularities of correlator in boundary spacetime coordinates is also a challenge for measurements. Another well-known method of reconstruction is called ``entanglement wedge reconstruction'', which uses the boundary entanglement spectrum to reconstruct bulk geometry and local observables~\cite{Dong:2016eik,Espindola:2018ozt,Penington:2019npb,Faulkner:2017vdd}. 
There are also some other approaches based on holographic entanglement entropy~\cite{Ryu:2006bv,Nishioka:2009un,Jokela:2020auu,Ahn:2024lkh} (furthermore, the work in ~\cite{Lashkari:2013koa} demonstrated that entanglement entropy can be utilized to derive the linearized Einstein equations governing the bulk metric.), Wilson loop~\cite{Maldacena:1998im,Rey:1998ik}, conformal (or Virasoro) symmetry~\cite{Nakayama:2015mva,Verlinde:2015qfa,Czech:2016xec,Anand:2017dav,Chen:2023naw},  tensor network~\cite{Swingle:2009bg,Pastawski:2015qua,Cao:2020uvb,Qi:2013caa,PhysRevLett.101.110501} or complexity~\cite{Susskind:2014rva,Susskind:2014jwa,Brown:2015bva}, e.g. see Refs.~\cite{Hashimoto:2020mrx,Hashimoto:2021umd,Xu:2023eof,Bilson:2010ff,Caputa:2017urj,Milsted:2018san}. In these approaches, some boundary quantities are conjectured to dual to some bulk geometric quantities and so give the chance to recover the bulk metric. For example, by considering the inverse problem of bulk minima surfaces enclosed at boundary, one may obtain the information of bulk metric~\cite{Alexakis_2020,Bao:2019bib}. It needs to be noted that except for a few of special systems (such a recent example for  glue Yang-Mills theory has been discussed in Ref.~\cite{Jokela:2023yun} ), it is still a challenge to experimentally measure the entanglement entropy or Wilson loop of general materials. The complexity currently is a ``theoretical quantity'' and the way to measure it is unclear. There have been related studies using machine learning to recover the bulk geometry of materials, see~\cite{Ahn:2024lkh,Ahn:2024gjf}. Ref.~\cite{Engelhardt:2016crc} points out the possibility to reconstruct the conformal metric in the bulk without needing to assume any bulk symmetries, by using lightcone singularities. However, considering that in the potential applications for real situations, it still interesting to ask if we have any simple method to reconstruct the exact bulk metric for the static, isotropic and homogeneous boundary.  In addition, for materials in real situations, we have ways to measure their 2-point correlation function in the laboratory. Considering the fact that 2-point function is earliest quantity appearing in the study of holography and one of most common measurable quantities in labs, we then ask a natural question: is there a universal and explicit method to reconstruct the bulk geometry by according to the 2-point functions of the dual materials?


This paper will give a positive answer. It will propose a new method about bulk reconstruction for a static asymptotically AdS black brane/hole by solving the inverse problem of 2-point correlation functions. This supplies a possibility to ``directly measure'' the corresponding holographic spacetime. The purpose of following parts is to present a proof of following theorem,
\begin{theorem}\label{scalarthm1}
For a stable planar symmetric static asymptotically AdS black brane, the bulk geometry can be reconstructed from the boundary frequency 2-point correlation functions of arbitrary two probe scalar operators that are dual to Klein-Gordon fields in the bulk.
\end{theorem}
It will also give explicit steps about how to reconstruct the bulk metric by using BTZ black hole as a concrete example. Here it uses the Klein-Gorden field to explain the main idea of this reconstruction. At the end of this paper, it will show that this requirement can be relaxed. It is important to note that our method requires the bulk should have a globally planar symmetry, so it may do not work on some spacetimes with complex structures (e.g. a bulk which is gluing of multiple distinct spacetimes).


\section{Overview on correlation function in holography}
Before we present our proof of the theorem above and the method of reconstructing bulk metric, let us first briefly review the ``solving direct problem'' of holographic 2-point correlation functions. We focus on the planar symmetric asymptotically AdS black brane, of which the metric will have following form
\begin{equation}\label{adsmetric}
  \td s^2=\frac{1}{z(\rho)^2}[h(\rho)(-\td t^2+\td\rho^2)+\td\vec{x}^2_{d-1}],~~\rho\in\mathbb{R}^+\,.
\end{equation}
The functions $h(\rho)$ and $z(\rho)$ will satisfy following boundary conditions
\begin{equation}\label{asymh1}
  z(0)=0,~~z'(0)=h(0)=1\,
\end{equation}
and
\begin{equation}\label{asymh2}
  z(\infty)=z_h,~~h(\infty)=0\,.
\end{equation}
The horizon then locates at $\rho\rightarrow\infty$ and the AdS boundary locates at $\rho=0$.

When we consider the correlation function of scalar operator in holography, we consider a real probe scalar field $\Psi$ which is described by Klein-Gorden equation $\nabla^2\Psi+m^2\Psi=0$ in the bulk. Here $m^2$ is the mass-square of the scalar field and is relative to the conformal dimension $\Delta$ of boundary scalar operator $\mathcal{O}$ such that $\Delta=d/2+\sqrt{d^2/4+m^2}$. In general, the asymptotical expansion of the scalar field near the AdS boundary has two independent branches,
\begin{equation}\label{aspseries1}
  \Psi=\Phi^{(+)}_\Delta \rho^{\Delta}(1+\cdots)+\Phi^{(-)}_\Delta \rho^{d-\Delta}(1+\cdots)\,.
\end{equation}
Here we choose that so called ``standard quantization'', which treats $\Phi^{(-)}_\Delta$ as the source term of boundary theory and $\Phi^{(+)}_\Delta\sim\langle\mathcal{O}\rangle$ as the value of expectation for boundary scalar operator.  In some cases, we may have an alternative quantization by choosing $\Phi^{(+)}_\Delta$ as the source, for which the conformal dimension of dual operator is $d-\Delta$. In the following discussion, we shall focus on the ``standard quantization''. The generalization to alternative quantization is parallel. The Weyl anomaly coming from gravity is zero~\cite{Henningson:1998ey,deHaro:2000vlm} due the flatness of boundary. The Weyl anomaly of the matter sector will appear for some fine-tuned values of mass-square, so we will do not consider it here.

To compute the frequency 2-point correlation function of dual boundary scalar operator, we consider the bulk scalar field of a form $\Psi=z^{(d-1)/2}\phi(\rho)\e^{-i\omega t}$ on the fixed background~\eqref{adsmetric}. The Klein-Gordon equation then reads
\begin{equation}\label{eqforscalar1}
  -\frac{\td^2\phi}{\td\rho^2}+\left(V_{\Delta}+\frac{4m^2+d^2-1}{4\rho^2}\right)\phi=\omega^2\phi\,.
\end{equation}
Here
\begin{equation}\label{defVeff}
\begin{split}
  V_{\Delta}&=-\frac{(d-1)z''}{2z}+\frac{(d^2-1)}4\left(\frac{z'^2}{z^2}-\frac1{\rho^2}\right)+m^2\left(\frac{h}{z^2}-\frac1{\rho^2}\right)\,.
  \end{split}
\end{equation}
%
Eq.~\eqref{eqforscalar1} has four kinds of solutions $\{\psi_\omega(\rho), \psi_\omega(\rho)^*, \varphi^{(+)}_\omega(\rho), \varphi^{(-)}_\omega(\rho)\}$. The $\{\psi_\omega(\rho), \psi_\omega(\rho)^*\}$ are defined by following boundary conditions
\begin{equation}\label{bdfors4}
  \lim_{\rho\rightarrow\infty}\e^{-i\omega\rho}\psi_\omega(\rho)=\lim_{\rho\rightarrow\infty}\e^{i\omega\rho}\psi_\omega(\rho)^*=1\,.
\end{equation}
The solutions $\{\varphi^{(\pm)}\}$ are defined by following boundary conditions when $\rho\rightarrow0^+$
\begin{equation}\label{varphipm1}
  \varphi^{(\pm)}_\omega(\rho)=\rho^{1/2\pm v}[1+\mathcal{O}(\omega^2\rho^2)]\,.
\end{equation}
Here $v=\sqrt{d^2/4+m^2}=\Delta-d/2$. Due to the fact that Eq.~\eqref{eqforscalar1} is 2nd order linear differential equation with real-valued coefficients, any one of above 4 types of solutions can be expressed in terms of linear combination of any other two. In the case that we consider the retarded Green function of holography, we impose the ingoing wave boundary condition $\Phi\propto \e^{-i\omega (t-\rho)}$ near the horizon. This leads to $\phi(\rho)\propto\psi_\omega(\rho)$. We then have
\begin{equation}\label{solvepsistar}
  \psi_{\omega}(\rho)=\Phi^{(-)}_{\Delta}(\omega)\varphi_{\omega}^{(-)}(\rho)+\Phi^{(+)}_{\Delta}(\omega)\varphi_{\omega}^{(+)}(\rho)\,.
\end{equation}
The frequency 2-point correlation function $\mathcal{G}(\omega)$ is defined as~\footnote{Here we omit an irrelevant factor that is determined by scaling dimension.}
\begin{equation}\label{defG1}
  \mathcal{G}_{\Delta}(\omega)=\Phi^{(+)}_{\Delta}(\omega)/\Phi^{(-)}_{\Delta}(\omega)\,.
\end{equation}
In the ``solving direct problem'' of correlation functions, what we have known is the metric functions $\{z(\rho),h(\rho)\}$ and the purpose is to find the correlation function $\mathcal{G}_\Delta(\omega)$ for a give conformal dimension $\Delta$.

For the purpose of proving Theorem~\ref{scalarthm1} and following notations in Refs.~\cite{Chadan1989,Kravchenko2020,Kravchenko2021}, let us introduce a function $F_{\Delta}(\omega)$ such that
\begin{equation}\label{deffw1}
  F_{\Delta}(\omega)=\lim_{\rho\rightarrow0^+}\frac{\sqrt{2\pi}(-\omega\rho)^{v-1/2}}{2^v\Gamma(v)}\e^{i\pi(v-1/2)/2}\psi_\omega(\rho)\,.
\end{equation}
This function plays an important role in reconstruction of potential $V_\Delta$, which is not determined by correlation function. However, this paper proves following lemma to relate it to correlation function,
\begin{lemma}\label{F2Gw}
The modulus of $F_{\Delta}(\omega)$ is completely determined by the correlation function
\begin{equation}\label{relF2C2}
  |F_{\Delta}(\omega)|^2=\frac{\pi|\omega/2|^{2v}v}{\Gamma(v+1)^2\Im[\mathcal{G}_{\Delta}(\omega)]}\,.
\end{equation}
\end{lemma}
See appendix A of supplemental materials for a proof. 

\section{ ``Solving the inverse problem'' and bulk reconstruction}
The ``solving the inverse problem'' of correlation functions then is to recover the metric functions $\{z(\rho),h(\rho)\}$ from the correlation $\mathcal{G}_\Delta(\omega)$. Though our reconstruction will be performed in real-valued frequency, let us make a short discussion on mathematical property of analytical continuation in complex frequency.

The function $\psi_\omega(\rho)$ is not square-integrable for a general $\omega\in\mathbb{C}$. However, there may be some special complex frequencies $\omega_k$ such that $\Phi^{(-)}(\omega_k)=0$. Such modes are called ``quasi-normal modes (QNMs)''. The QNMs are related to the stability of the holographic system. Particularly, if there is a QNM $\omega_k$ such that $\Im(\omega_k)>0$, then the system will be unstable under perturbations since the scalar field $\Psi=z^{(d-1)/2}\phi(\rho)\e^{-i\omega_k t}\propto\e^{\Im(\omega_k)t}$ will grow exponentially. Instead, the system will be stable if no QNM has positive imaginary part. For those unstable QNMs we have following lemma
\begin{lemma}\label{thm5}
If $\omega_k\in\mathbb{C}$ is an unstable QNM, then $\psi_{\omega_k}$ is square-integrable in the interval $(0,\infty)$. If $\psi_{\omega_k}$ is square-integrable in the interval $(0,\infty)$, then $\omega_k$ is a QNM and $\omega_k^2<0$.
\end{lemma}
A proof will be found in appendix B of supplemental materials.

From this lemma we see that the system is stable if and only if there is no square-integrable solution for Eq.~\eqref{eqforscalar1}. In practice, the correlation function is measured according to linear response of the system to external perturbations. Then the system should be stable against the small perturbations otherwise the correlation function can not be obtained experimentally. Thus, it is natural to assume that Eq.~\eqref{eqforscalar1} \textit{has no square-integrable solution}. Under such an assumption, we will have following mathematical results to reconstruct effective potential (see the chapters 1-3 of Ref.~\cite{Chadan1989} for more details and explanations on its mathematical aspects)
\begin{equation}\label{effVK1}
  V_{\Delta}(\rho)=2\frac{\td}{\td \rho}K_{\Delta}(\rho,\rho)\,.
\end{equation}
The transmutation kernel $K_{\Delta}(\rho, \sigma)$ is determined by the Gel'fand-Levitan-Marchenko integral equation
\begin{equation}\label{GLMeqs}
  K_{\Delta}(\rho,\sigma)+\Omega_{\Delta}(\rho,\sigma)+\int_0^\rho K_{\Delta}(\rho,y)\Omega_{\Delta}(y,\sigma)\td y=0
\end{equation}
and function $\Omega_{\Delta}(\rho,\sigma)$ is given by function $F_{\Delta}(\omega)$ according to
\begin{equation}\label{defOmegav}
  \Omega_{\Delta}(\rho,\sigma)=\int_0^{\infty}\hat{J}_{v}(\omega\rho)\hat{J}_{v}(\omega\sigma)\left(\frac1{|F_{\Delta}(\omega)|^2}-1\right)\td\omega\,.
\end{equation}
Here $\hat{J}_v(x)=\sqrt{x}J_v(x)$ and $J_{v}(x)$ is the $v$-th order Bessel function of the first kind. When the system is not stable, i.e. correlation function has poles at upper complex plane, we still have similar results but the function $\Omega_{\Delta}(\rho,\sigma)$ will be more complicated, see Refs.~\cite{Chadan1989,Kravchenko2020,Kravchenko2021}. According to Lemma~\ref{F2Gw}, the effective potential can be recovered from the correlation function.

Take $\mathcal{G}_{\Delta_1}(\omega)$ and $\mathcal{G}_{\Delta_2}(\omega)$ to be correlation functions of arbitrary two different real scalar operators. After we reconstruct their corresponding effective potentials $V_{\Delta_1}$ and $V_{\Delta_2}$, the function $z(\rho)$ will be determined by
\begin{equation}\label{diff2zrho}
\begin{split}
  \frac{m_2^2V_{\Delta_1}-m_1^2V_{\Delta_2}}{m_2^2-m_1^2}&=\frac{(d^2-1)}4\left(\frac{z'^2}{z^2}-\frac1{\rho^2}\right)-\frac{(d-1)z''}{2z}\,.
  \end{split}
\end{equation}
The function $h(\rho)$ then is given by
\begin{equation}\label{solvehrho}
  h=\frac{V_{\Delta_1}-V_{\Delta_2}}{m_1^2-m_2^2}z^2+\frac{z^2}{\rho^2}\,.
\end{equation}
Naively, one may think that we can solve Eq.~\eqref{diff2zrho} by using initial condition~\eqref{asymh1}, i.e. $\{z(0)=0,z'(0)=1\}$. However, such a Cauchy problem is ill-defined due to that the Eq.~\eqref{diff2zrho} is singular at $z=0$. From the physical viewpoint, this is because that the solution of above system is not unique but can be different up to a scaling transformation. Instead of solving Eq.~\eqref{diff2zrho} as a Cauchy problem, we solve this problem as a boundary-value-problem by requiring
\begin{equation}\label{boundvalue}
  \lim_{\rho\rightarrow0^+}z/\rho=1,~~\text{and}~z(\infty)=z_h\,.
\end{equation}
The first boundary condition insures metric is asymptotically AdS when $\rho\rightarrow0^+$. Without losing generality, we can set ``inverse horizon radius'' $z_h=1$ by using the scaling transformation. This setup fixes the scaling symmetry. The boundary-value-problem now is well defined and so we finished the proof of Theorem~\ref{scalarthm1}.

\textit{Example of BTZ black hole}-.
Let us consider the BTZ black hole as an example and exhibit how to recover the metric from its 2-point correlation functions. By setting the inverse horizon $z_h=1$, the BTZ black hole is given by $z(\rho)=\tanh\rho$ and $h(\rho)=1/\cosh^2\rho$. The boundary 2-point correlation function is give by
\begin{equation}\label{btzcorrel1}
  \mathcal{G}_{\Delta}(\omega)=\frac{\Gamma(-v)\Gamma(\frac{1-i\omega+v}2)^2}{\Gamma(v)\Gamma(\frac{1-i\omega-v}2)^2}\,.
\end{equation}
The correlation function has pole when $\omega_k=-i(1+v+2k)$ with $k\in\mathbb{N}^+$. Thus there is no QNM at upper complex plane and so no unstable QNM.

In following we will applied the ``Fourier-Jacobi series representation''  to solve Eq.~\eqref{GLMeqs}. Since it may not be familiar to the potential readers of this paper, we here first briefly explain this method. For more mathematical aspects, one can refer to Refs.~\cite{Kravchenko2020,Kravchenko2021} and references therein.

\textit{Step 1}) We write the kernel into following series expansion
\begin{equation}\label{seriesK1}
  K_{\Delta}(\rho,\sigma)=\sum_{n=0}^\infty\frac{\beta_n(\rho)}{\rho^{v+3/2}}\sigma^{v+1/2}P_n^{(v,0)}(1-2\sigma^2/\rho^2)\,,
\end{equation}
where $P_n^{(\alpha,\beta)}$ stands for the Jacobi polynomial. Introduce a matrix $A_{ln}$ and an vector $B_n$ such that
\begin{equation}\label{defAmn}
  A_{ln}(\rho)=\int_0^{\infty}\frac{J_{v+1+2l}(\omega\rho)J_{v+1+2n}(\omega\rho)}{\omega\rho}\left(\frac1{|F_{\Delta}(\omega)|^2}-1\right)\td\omega
\end{equation}
and
\begin{equation}\label{defBm}
  B_n(\rho)=-\int_0^{\infty}J_{v}(\omega\rho)J_{v+1+2n}(\omega\rho)\left(\frac1{|F_{\Delta}(\omega)|^2}-1\right)\td\omega\,.
\end{equation}
Then Eq.~\eqref{GLMeqs} becomes a discrete form
\begin{equation}\label{discreteGLM}
  \frac{\beta_l(\rho)}{2(2l+v+1)\rho}+\sum_{n=0}^{M} A_{ln}(\rho)\beta_n(\rho)=B_l(\rho)\,.
\end{equation}
Here $l=0,1,2\cdots,M$ with $M\rightarrow\infty$. To perform numerical computation, we can truncate system into a finite system by setting a finite $M$. \\
\textit{Step 2}) Discrete the coordinate $\rho$ to be $\rho_k=k\delta\rho$. For every $\rho_k$, compute $B_m(\rho_k)$ and $A_{mn}(\rho_k)$ according to the formulas~\eqref{defAmn} and \eqref{defBm}. One then solves the truncated system \eqref{discreteGLM} to find $\beta_0(\rho_k)$ for every $\rho_k$.\\
\textit{Step 3}) Use the finite differential methods to obtain the $\beta'_0(\rho_k)$ and $\beta''_0(\rho_k)$ based on the values of $\beta_0(\rho_k)$. The effective potential at the $k$-th grid can be recovered according to
\begin{equation}\label{effVbeta0}
  V_{\Delta}(\rho_k)=\frac{\rho\beta_0''(\rho_k)+(2v+1)\beta'_0(\rho_k)}{\rho_k[\beta_0(\rho_k)+2(v+1)]}\,.
\end{equation}
\textit{Step 4}) Repeat the steps (1)-(3) to obtain effective potentials $\{V_{\Delta_1}, V_{\Delta_2}\}$ for two different scaling dimension and then we use spline interpolation to obtain two continuous functions for $V_{\Delta_1}, V_{\Delta_2}$ and solve the differential equation Eq.~\eqref{diff2zrho} to obtain function $z(\rho)$.\\
\textit{Step 5}) Take the solution $z(\rho_k)$ into Eq.~\eqref{solvehrho} to obtain the function $h(\rho_k)$.

\begin{figure}
	\centering
	\includegraphics[width=0.75\textwidth]{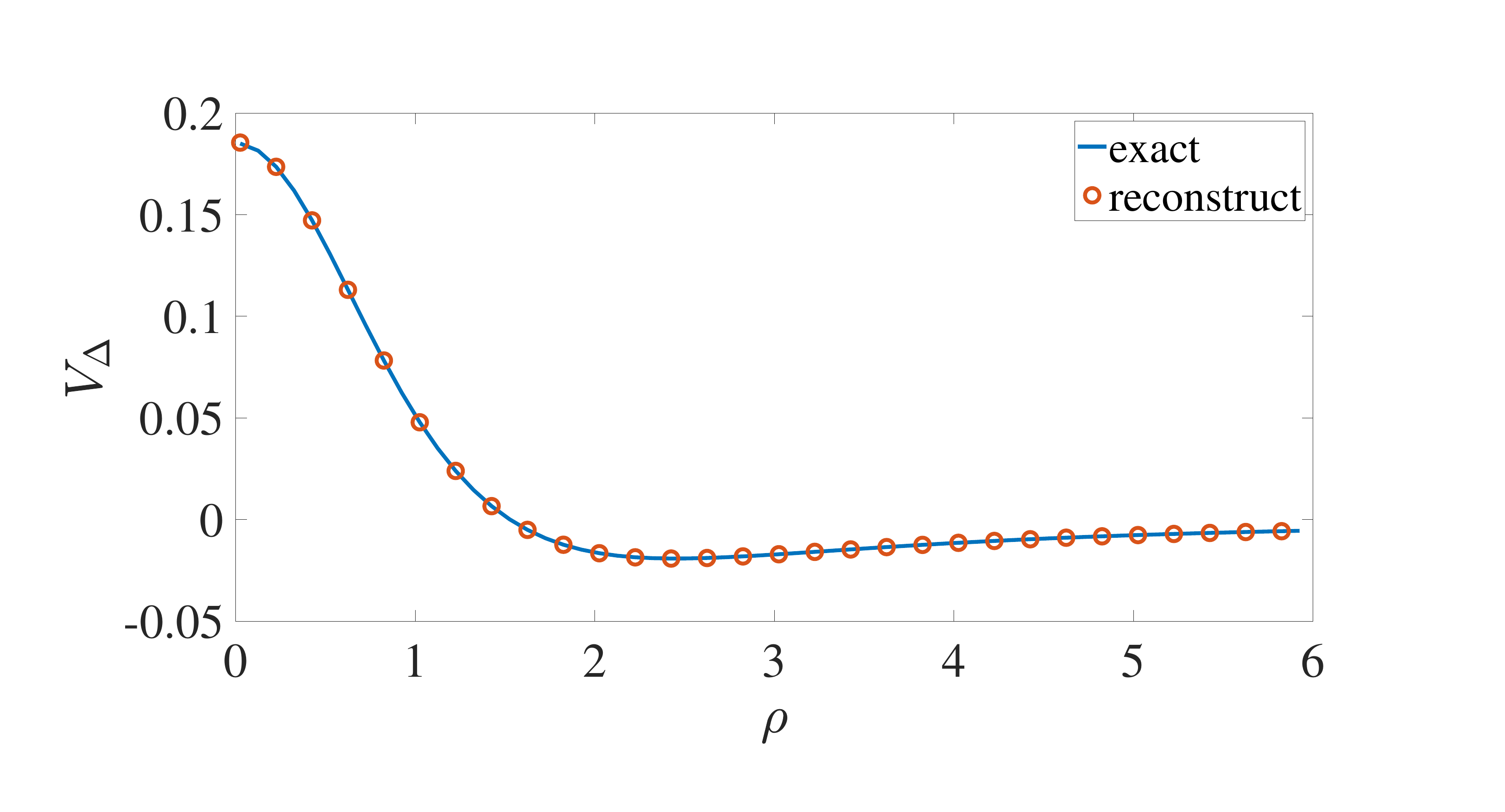}
	\includegraphics[width=0.75\textwidth]{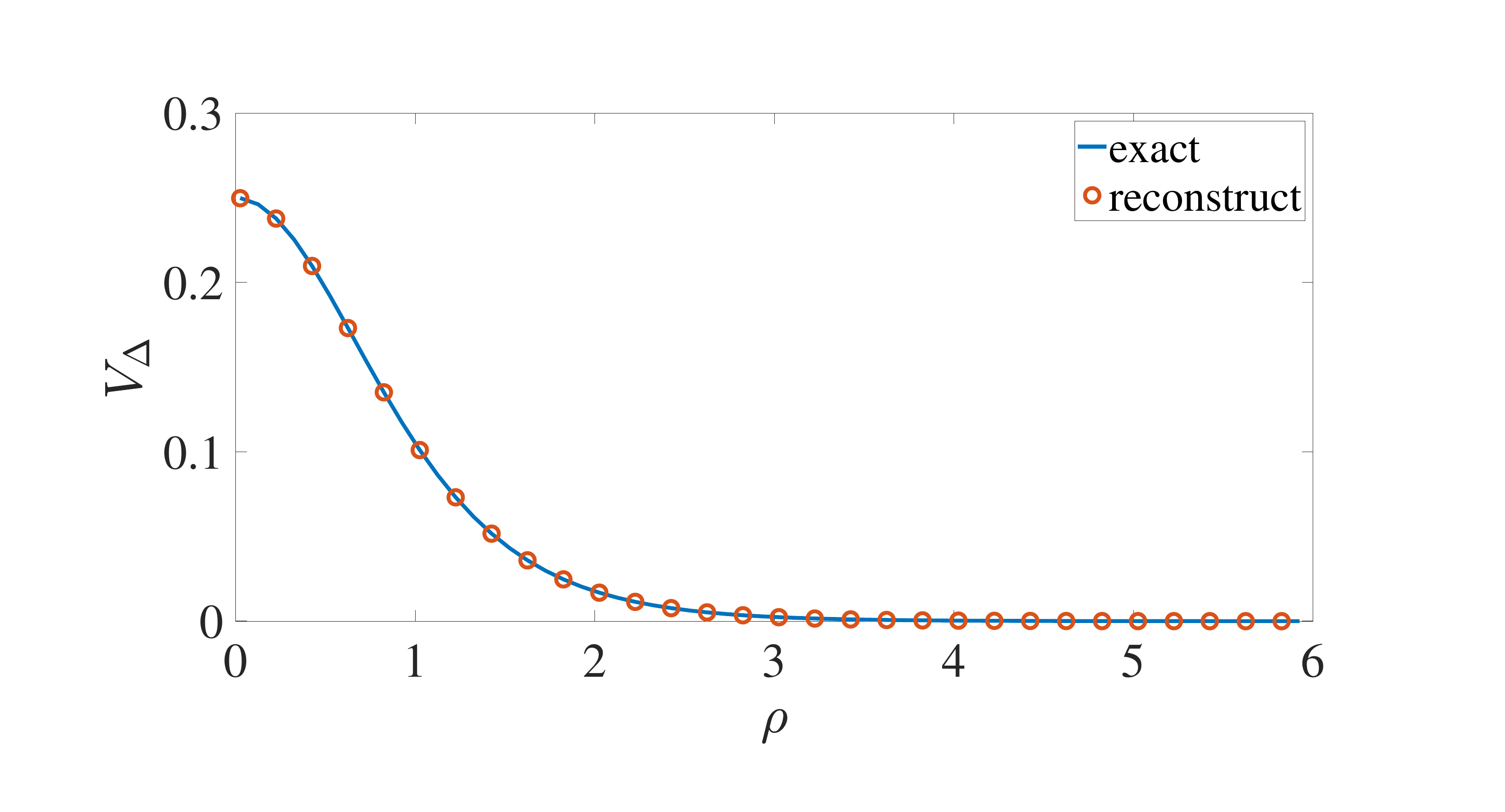}
	\caption{Comparison between the reconstructed potential and their exact values by using 5 equations in the truncated system. Here it uses correlation functions of $\Delta=5/3$ (above) and $\Delta=3/2$ (below) to recover the potentials $V_{\Delta}$.}\label{potential}
\end{figure}

\begin{figure}
\centering
    \includegraphics[width=0.55\textwidth]{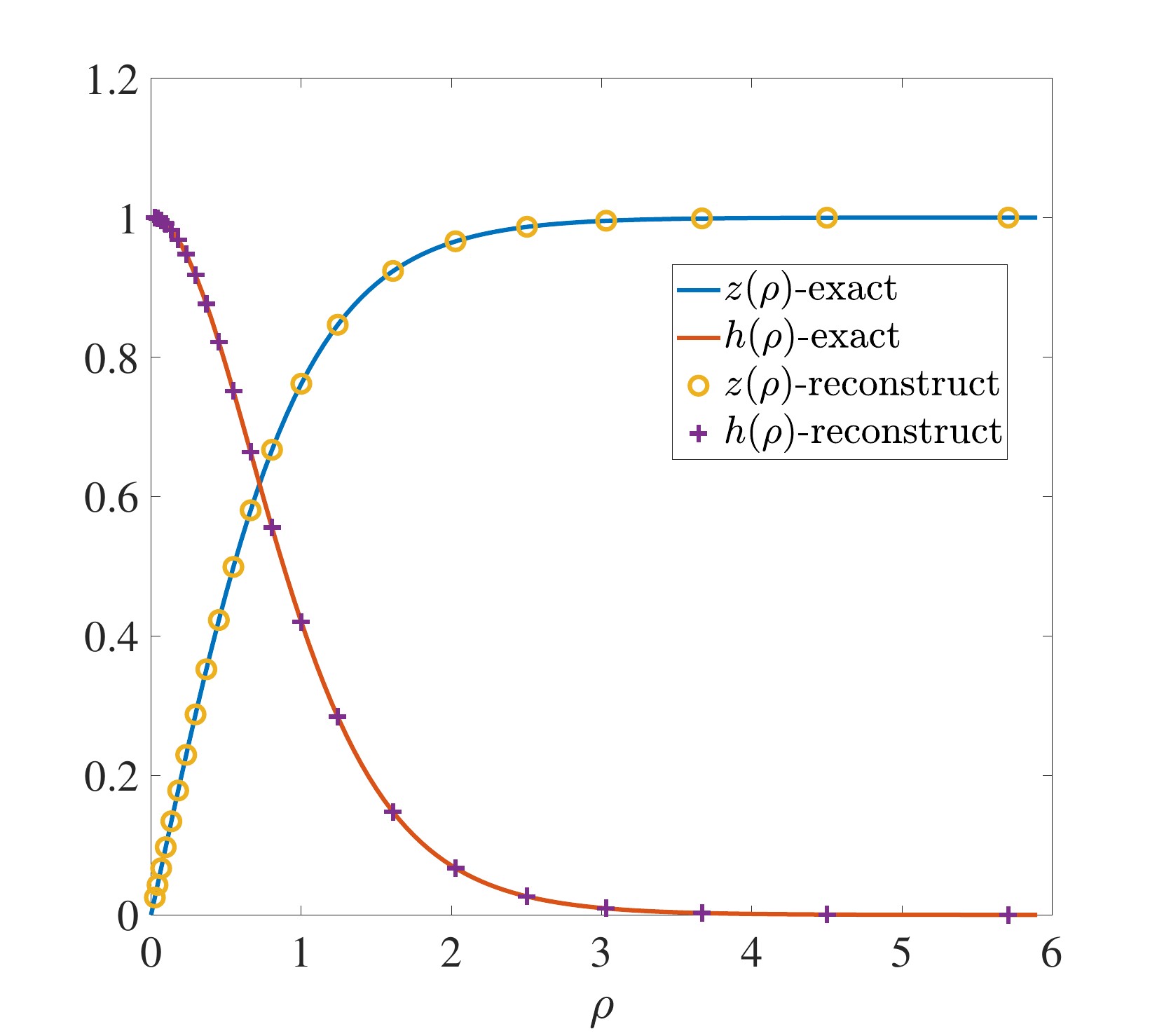}
   \caption{Comparison between the reconstructed metric components and their exact values by using 5 equations in the truncated system. Here it uses correlation functions of $\Delta=5/3$ and $\Delta=3/2$ to recover the metric.}\label{fighzbtz}
\end{figure}

In Fig.~\ref{potential} and Fig.~\ref{fighzbtz} we show the reconstructed potentials and metric components by using our numerical method. Here we choose the scaling dimensions of dual boundary theory to be $\Delta=5/3$ and $\Delta=3/2$ and only use 5 truncated equations (i.e. setting $M=4$). We see that the reconstructed method gives excellent results.
%
%

\section{Discussion}
The requirement that bulk field satisfies the Klein-Gorden equation can be relaxed. Consider a general Lagrangian of scalar field $L=-\frac12[g^{\mu\nu}\partial_\mu\Psi\partial_\nu\Psi+m^2\Psi^2+W(\{X_i\},R_{\mu\nu\sigma\tau},g_{\mu\nu},\Psi)]$. Here the potential $W$ of scalar field may depend on curvature tensor $R_{\mu\nu\sigma\tau}$, bulk metric $g_{\mu\nu}$ and some others bulk matters fields $\{X_i\}$.  To have a well-defined linear-response theory, $W$ is required to satisfy $\partial W/\partial\Psi|_{\Psi=0}=0$. In this case, the linear-response theory in holography is given by tuning on an infinitesimal scalar source and Eq.~\eqref{eqforscalar1} becomes
\begin{equation}\label{eqforscalar3}
  -\frac{\td^2\phi}{\td\rho^2}+\left(V_{\Delta}+\frac{4m^2+d^2-1}{4\rho^2}+\tilde{V}\right)\phi=\omega^2\phi\,.
\end{equation}
with
\begin{equation}\label{tildeV1}
  \tilde{V}=\frac{h}{z^2}\left.\frac{\partial^2 W}{\partial\Psi^2}\right|_{\Psi=0}\,.
\end{equation}
In such a case, we can use two such operators of different $m^2$ to reconstruct the metric components once the function $W$ is known. Though the Theorem~\ref{scalarthm1} and above discussion focused on real scalar field and planar symmetric case, the generalizations to other kinds of operators and spherical/hyperbolic symmetric case are straightforward.

From the above discussions, one can realize that the boundary 2-point correlation functions of fields that only couple with gravity are not independent. We can use only of few of them to reconstruct the bulk metric and all the others will be determined by bulk metric. In addition, the entanglement entropy of large-$c$ limit in holography is completely determined by bulk metric~\cite{Ryu:2006bv,Nishioka:2009un}, so it can also be determined by the boundary 2-point correlation functions. On the other hand, one can use different operators to probe the material. To have a holographic dual for this material, the metrics reconstructed from these different correlation functions must be the same. The result in this paper then not only proposes an explicit method to build a holographic spacetime for a material that has holographic dual but also provides a powerful approach to check if a material has holographic dual. In addition, we notice that Refs.~\cite{Harlow:2018fse,Goldar:2024crc} suggests that recovering the entire bulk operator of a free scalar field (see the
 Eq. (3.10) of Ref.~\cite{Harlow:2018fse}) requires the following 3 points: (1) dynamics of probe field, i.e. the action of free scalar field, (2) the bulk metric so that it can obtain the bulk  smearing function $K$, and (3) boundary data of dual scalar theory.  Our work reduces this requirement because we found that the bulk metric can be recovered from the boundary data at least for free scalar field. The reconstruction proposed here requires that the dynamics of probe field is known. It needs to note that the probe field can be chosen by experimenters and we can choose the known probe fields. In addition, it not very hard to realize that above method can be
 generalized into fermion fields, vector fields, $p-$form fields or other complicated fields.

In fact, the correlation functions belong to the scattering matrix of bulk theory. For the case of less symmetries, it is unclear about how to recover the bulk geometry from its scattering matrix. However, if we assume that the metric is analytical except for intrinsic singularities, the results of  Refs.~\cite{Barreto,Joshi2000,doi:10.1142/e040,Hora2015,10.1215/S0012-7094-04-12911-2} show that the Euclidean asymptotically AdS manifold and its metric, up to invariants, are determined by the scattering matrix at all energies. This implies, at least the general static asymptotically AdS black brane/hole (only in this case Wick rotation to Euclidean signature is unambiguous) is determined by its boundary correlation functions~\footnote{Refs.~\cite{Barreto,Joshi2000,doi:10.1142/e040,Hora2015} discuss the case that $v$ is pure imaginary number. However, the scattering matrix with respective to $v$ has a meromorphic continuation to the complex plane~\cite{10.1215/S0012-7094-04-12911-2}. }~\footnote{The author thanks Prof. Ant\^{o}nio S\'{a} Barreto for telling me the existence of meromorphic continuation by email. }, though the explicit method of reconstruction is still unknown. This is still an active topic (see Ref.~\cite{Isozaki2020} for a recent review on inverse scattering problem), and it hopes that some useful progresses will be made in the future.

\begin{acknowledgments}
This work is supported by the Natural Science Foundation of China under Grant No. 12375051.
\end{acknowledgments}

\appendix
\section{Proof of Lemma 1}
Apart from Eq.~\eqref{solvepsistar}, we can also expand solutions $\varphi_{\omega}^{(\pm)}(\rho)$ in terms of the combinations of $\{\psi_{\omega}(\rho), \psi_{\omega}(\rho)^*\}$
\begin{equation}\label{expanpsi10}
	\varphi_{\omega}^{(+)}(\rho)=\frac{i}{2\omega}\left[C_{\Delta}^{(+)}(\omega)\psi_{\omega}(\rho)^*-C_{\Delta}^{(+)*}(\omega)\psi_{\omega}(\rho)\right]\,
\end{equation}
and
\begin{equation}\label{expanpsi20}
	\varphi_{\omega}^{(-)}(\rho)=\frac{1}{2\omega}\left[C_{\Delta}^{(-)}(\omega)\psi_{\omega}(\rho)^*+C_{\Delta}^{(-)*}(\omega)\psi_{\omega}(\rho)\right]\,.
\end{equation}
From boundary conditions~\eqref{bdfors4} we can find
\begin{equation}\label{phipsiri}
	\psi_\omega(\rho)^*=\psi_{-\omega}(\rho)\,.
\end{equation}
for $\omega\in\mathbb{R}$. From the fact that Eq.~\eqref{eqforscalar1} is invariant under transformation $\omega\rightarrow-\omega$, we will see that
\begin{equation}\label{evenvarphi0}
	\varphi^{(\pm)}_\omega(\rho)=\varphi^{(\pm)}_{-\omega}(\rho)=\varphi^{(\pm)}_\omega(\rho)^*\,
\end{equation}
for $\omega\in\mathbb{R}$.

We can solve the $\psi_{\omega}$ from Eqs.~\eqref{expanpsi10} and \eqref{expanpsi20}. Take Eq.~\eqref{solvepsistar} into account and we then find
\begin{equation}\label{cvaluephi2}
	\begin{split}
		\Phi^{(+)}_{\Delta}&=\frac{-2i\omega C_{\Delta}^{(-)}}{C_{\Delta}^{(-)}C_{\Delta}^{(+)*}+C_{\Delta}^{(-)*}C_{\Delta}^{(+)}}\\
		\Phi^{(-)}_{\Delta}&=\frac{2\omega C_{\Delta}^{(+)}}{C_{\Delta}^{(-)}C_{\Delta}^{(+)*}+C_{\Delta}^{(-)*}C_{\Delta}^{(+)}}\,.
	\end{split}
\end{equation}
The correlation function $\mathcal{G}(\omega)$ then is given by
\begin{equation}\label{defG1}
	\mathcal{G}_{\Delta}(\omega)=\frac{\Phi^{(+)}_{\Delta}(\omega)}{\Phi^{(-)}_{\Delta}(\omega)}=-i\frac{C_{\Delta}^{(-)}(\omega)}{C_{\Delta}^{(+)}(\omega)}\,.
\end{equation}

Let us define  Wronskian determinate $W[f,g]=fg'-f'g$. For any two solutions of Eq.~\eqref{eqforscalar1}, their Wronskian is constant. We now compute $W[\psi_\omega,\psi_{-\omega}]$ at the infinity and $W[\varphi^{(+)}_\omega,\varphi^{(-)}_\omega]$ at AdS boundary. The results then read
\begin{equation}\label{wron1}
	W[\psi_\omega,\psi_{-\omega}]=-2i\omega,~~W[\varphi^{(+)}_\omega,\varphi^{(-)}_\omega]=-2v\,.
\end{equation}
The function $C_{\Delta}^{(\pm)}(\omega)$ can also be determined according to Wronskian determinate
\begin{equation}\label{deffl1}
	C_{\Delta}^{(+)}(\omega)=W[\psi_{\omega},\varphi_\omega^{(+)}],~~C_{\Delta}^{(-)}(\omega)=iW[\psi_{\omega},\varphi_\omega^{(-)}]\,.
\end{equation}
These can be verified by using Eqs.~\eqref{expanpsi10}, \eqref{expanpsi20} and \eqref{wron1}. Using the asymptotically behavior Eq.~\eqref{varphipm1} and computing Eq.~\eqref{deffl1} near $\rho=0$, we can find
\begin{equation}\label{valuecvs1}
	C_{\Delta}^{(+)}(\omega)=2v\lim_{\rho\rightarrow0^+}\rho^{v-1/2}\psi_\omega(\rho)
\end{equation}
if $v>0.$

We note from Eqs.~\eqref{expanpsi10} and \eqref{expanpsi20} that
\begin{equation}\label{wvarphi12s1}
	\begin{split}
		W[\varphi^{(+)}_\omega,\varphi^{(-)}_\omega]=\frac{i}{4\omega^2} W[\psi_{-\omega},\psi_{\omega}]\left(C_{\Delta}^{(-)}(\omega)C_{\Delta}^{(+)}(\omega)^*+C_{\Delta}^{(-)}(\omega)^*C_{\Delta}^{(+)}(\omega)\right)\,.
	\end{split}
\end{equation}
Thus, we obtain
\begin{equation}\label{eqforcc1}
	C_{\Delta}^{(-)}(\omega)C_{\Delta}^{(+)}(\omega)^*+C_{\Delta}^{(-)}(\omega)^*C_{\Delta}^{(+)}(\omega)=4v\omega\,.
\end{equation}
On the other hand, Eq.~\eqref{defG1} shows
\begin{equation}\label{cpmgw}
	C_{\Delta}^{(-)}(\omega)=i\mathcal{G}(\omega)^*C_{\Delta}^{(+)}(\omega)\,.
\end{equation}
Take it into Eq.~\eqref{eqforcc1} and we will find
\begin{equation}\label{absCpm1}
	|C_{\Delta}^{(+)}(\omega)|^2=2v\omega/\Im[\mathcal{G}_{\Delta}(\omega)]\,.
\end{equation}
The function $F_{\Delta}(\omega)$ is defined as
\begin{equation}\label{deffw10}
	F_{\Delta}(\omega)=\lim_{\rho\rightarrow0^+}\frac{\sqrt{2\pi}(-\omega\rho)^{v-1/2}}{2^v\Gamma(v)}\e^{i\pi(v-1/2)/2}\psi_\omega(\rho)\,.
\end{equation}
Then we see that the function $F_{\Delta}(\omega)$ will is related to the coefficient $C^{(+)}_{\Delta}$ such that
\begin{equation}\label{relF2C0}
	F_{\Delta}(\omega)=\frac{\sqrt{\pi}(-\omega/2)^{v-1/2}}{2\Gamma(v+1)}\e^{i\pi(v-1/2)/2}C^{(+)}_{\Delta}\,.
\end{equation}
so we find the modulus of $|F_{\Delta}(\omega)|$ is completely determined by the correlation function
\begin{equation}\label{relF2C20}
	|F_{\Delta}(\omega)|^2=\frac{\pi|\omega/2|^{2v}v}{\Gamma(v+1)^2\Im[\mathcal{G}_{\Delta}(\omega)]}\,.
\end{equation}
$\square$
\section{Proof of Lemma 2}
Since $\Im(\omega_k)>0$, we then find that $\psi_{\omega_k}$ will decay exponentially when $\rho\rightarrow\infty$. On the other hand, $\Phi^{(-)}(\omega)=0$ leads that $\psi_{\omega_k}\rightarrow0$ when $\rho\rightarrow0$. Thus we can conclude that $\psi_{\omega_k}$ is square-integrable in the interval $(0,\infty)$. On the other hand, if $\psi_{\omega_k}$ is square-integrable in the interval $(0,\infty)$,  we first should have $\lim_{\rho\rightarrow\infty}\psi_{\omega_k}=\e^{i\omega_k\rho}\rightarrow0$ and $\Phi^{(-)}=0$, so $\omega_k$ is a QNM. In addition, under these two boundary conditions, Eq.~\eqref{eqforscalar1} in fact forms a ``Sturm-Liouville eigenvalue problem'' and the $\omega^2_k$ is one of eigenvalue, so the $\omega^2_k$ must be real. The $\lim_{\rho\rightarrow\infty}\psi_{\omega_k}=\e^{i\omega_k\rho}\rightarrow0$  means  $\omega_k^2<0$.  $\square$
\bibliographystyle{JHEP}
\bibliography{inv-correl2}

\end{document}